\documentclass[showpacs,twocolumn,prl,floatfix,superscriptaddress]{revtex4}

\usepackage{times,amsmath,amsfonts,amssymb,latexsym}
\usepackage{graphicx,epsf}

\newcommand{\bra}[1]{\langle #1 |}
\newcommand{\ket}[1]{| #1 \rangle}

\newcommand\h{{\cal H}}

\newcommand{\be}{\begin{equation}}
\newcommand{\ee}{\end{equation}}
\newcommand{\ba}{\begin{eqnarray}}
\newcommand{\ea}{\end{eqnarray}}
\newcommand\tr{{\mbox{Tr\,}}}
\newcommand{\ignore}[1]{}
\def\CC{{\rm\kern.24em \vrule width.04em height1.46ex depth-.07ex
    \kern-.30em C}}
\def\P{{\rm I\kern-.25em P}}
\def\RR{{\rm
         \vrule width.04em height1.58ex depth-.0ex
         \kern-.04em R}}
\def\bbbone{{\mathchoice {\rm 1\mskip-4mu l} {\rm 1\mskip-4mu l}
{\rm 1\mskip-4.5mu l} {\rm 1\mskip-5mu l}}}
\def\bbbc{{\mathchoice {\setbox0=\hbox{$\displaystyle\rm C$}\hbox{\hbox
to0pt{\kern0.4\wd0\vrule height0.9\ht0\hss}\box0}}
{\setbox0=\hbox{$\textstyle\rm C$}\hbox{\hbox
to0pt{\kern0.4\wd0\vrule height0.9\ht0\hss}\box0}}
{\setbox0=\hbox{$\scriptstyle\rm C$}\hbox{\hbox
to0pt{\kern0.4\wd0\vrule height0.9\ht0\hss}\box0}}
{\setbox0=\hbox{$\scriptscriptstyle\rm C$}\hbox{\hbox
to0pt{\kern0.4\wd0\vrule height0.9\ht0\hss}\box0}}}}
\def\bbbz{{\mathchoice {\hbox{$\sf\textstyle Z\kern-0.4em Z$}}
{\hbox{$\sf\textstyle Z\kern-0.4em Z$}}
{\hbox{$\sf\scriptstyle Z\kern-0.3em Z$}}
{\hbox{$\sf\scriptscriptstyle Z\kern-0.2em Z$}}}}

\begin{document}
\title{Entanglement and area law with a fractal boundary in a
  topologically ordered phase}

\author{Alioscia Hamma}
%\email{ahamma@perimeterinstitute.ca}
\affiliation{Perimeter Institute for Theoretical Physics,
31 Caroline St. N, N2L 2Y5, Waterloo ON, Canada}
\author{Daniel A. Lidar}
\affiliation{Departments of Chemistry, Electrical Engineering, and Physics, and Center for Quantum Information Science and Technology, University of Southern California, Los Angeles CA 90089}
\author{Simone Severini}
\affiliation{Institute for Quantum Computing and Department of Combinatorics \& Optimization\\
University of Waterloo, 200 University Av. W, N2L 3G1, Waterloo ON,
Canada}

\begin{abstract}
Quantum systems with short range interactions are known to respect an
area law for the entanglement entropy: the von Neumann entropy $S$
associated to a bipartition scales with the boundary $p$ between the
two parts. Here we study the case in which the boundary
is a fractal. We consider the topologically ordered phase of the toric
code with a magnetic field. When the field vanishes it is possible
to analytically compute the entanglement entropy for both regular and
fractal bipartitions $(A,B)$ of the system, and this yields an upper
bound for the entire topological phase. When the $A$-$B$ boundary is
regular we have $S/p =1$ for large $p$.  When the boundary is a fractal of
Hausdorff dimension $D$, we show that the entanglement between the two
parts scales as
$S/p=\gamma\leq1/D$, and $\gamma$  depends on the fractal 
considered.
\end{abstract}


\maketitle

%%%%%%%%%%%%%%%%%%%%%%%%%%%%%%%%%%%%%%%%%%%%%%%%%%%%%%%%%%%

\textit{Introduction}.--- Entanglement is certainly one of the most
striking aspects of quantum theory. Not only is it the key ingredient
for protocols ranging from quantum teleportation to cryptography, but
it also has an important role in the study of condensed matter and
many body systems \cite{ent-review}. Quantum phase transitions can be
understood in terms of entanglement \cite{entqpt}, and new exotic
states of matter that defy a description in terms of local order
parameters show a signature of topological order in the global pattern
of their entanglement \cite{hiz,topentropy}. Moreover, the
analysis of the scaling of entanglement in the ground state of
condensed matter systems has shed new light on the question of their
simulability \cite{vidal}. 

Especially for the last reason, one is interested in knowing how
entanglement scales with the size of the system. If there is a gap, all correlations decay exponentially with the distance
in units of the length scale \cite{hastings-gap}. In this case, one also expects the entanglement
to be short ranged, so that only the degrees of freedom of the
boundary of the system contribute to the total entanglement. This is
the so called {\em area law} for the entanglement (see
Ref.~\cite{arealawrev} for a comprehensive review).

In this work, we study the case of a topologically ordered state, the
ground state of the toric code \cite{kitaev}. For this state -- and a
class of topologically ordered states -- the entanglement can be
computed \emph{exactly} \cite{hiz}.
For a bipartition with a regular boundary $p$, the entanglement measured by the von
Neumann entropy $S$ is exactly $S = p-1$, where the correction $-1$ is
due to a topological contribution to the entanglement \cite{hiz,topentropy}. Obviously,
$\gamma:= S/ p$ is $1$ in the
limit of large $p$. If we add perturbations to the model, topological
order is not destroyed until a quantum phase transition
happens. Throughout the entire topological phase the entanglement is
upper-bounded by its value in the unperturbed model \cite{topqpt-num}.

Here we study the case in which the boundary of
the system is a fractal curve of Hausdorff dimension $D$. This
situation arises under a large variety of experimental conditions in
two-dimensional systems \cite{lidarfractal}. The scaling of
entanglement for self similar systems is important also in view of
devising efficient algorithms
which use the renormalization group for
computing ground states of quantum systems in two dimensions
\cite{vidal}. One could expect that as the boundary of the system
becomes less regular, the entanglement increases with the length $p$
of the boundary, as in the case of fermions  \cite{fermions}.
In contrast to the fermionic case, we find that for
topologically ordered
spin systems the entanglement decreases with $p$. 
The length of a fractal curve -- and consequently the entanglement -- diverges in the limit of exact
fractality \cite{fractalbook}.
However, for every step $n$ of the iteration of the fractal, the
length of the curve is a finite number $p(n)$, which increases with
$n$.
In contrast to regular boundaries, for fractal
boundaries
$\gamma$ is a fractional number: we
can speak of \emph{fractal entanglement}. Moreover, we shall see that
$\gamma\leq D^{-1}$.

\textit{Entanglement and topological order}.--- Consider a unitary representation of a group $G$ acting on spin-$1/2$ degrees of freedom with Hilbert space $\h$. Since we wish to compute
the entanglement entropy associated to a bipartition of the system, we
are interested in the properties of the group when we split the
Hilbert space as $\h = \h_A \otimes \h_B$. We assume that there exists
a product state $\ket{0}= \ket{0_A} \otimes \ket{0_B}\in \h$. We can
now define the (normalized) {\em $G$-state} as 
$\ket{\Psi_G}:= \sum_{g\in G} \alpha(g) g \ket{0}$. If all the
coefficients are equal, we call the state a {\em $G$-uniform state}: 
$\ket{G}:= |G|^{-1/2} \sum_{g\in G} g \ket{0}$,
where $|G|$ is the order of $G$.
Note that
$\ket{G}$ is stabilized by the group $G$.
Let us now define the two subgroups of $G$ that act
trivially  on the subsystems $A,B$ respectively: 
$G_A := \{g\in G\ |\ \  g= g_A\otimes \bbbone_B\}$ and similarly for $G_B$. By defining the quotient group
$G_{AB}:=G/(G_A\times G_B)$, we can write $G$ as the union over all elements of $G_{AB}$:
$G= \bigcup_{[h]\in G_{AB}} \{ (g_A\otimes g_B)h\, | \ g_A\otimes
\bbbone_B \in G_A,\, \bbbone_A\otimes g_B\in G_B \}$. The state can thus be written as $\ket{\Psi_G} = |G|^{-1/2} \sum_{\substack{g_A\otimes g_B \in 
G_A\times  G_B \\ h\in G_{AB}}} \alpha (g_A\otimes g_B, h) h_A\otimes h_B\otimes (g_A \otimes g_B) \ket{0}$. If the
coefficients $\alpha$ in the expression for $\ket{\Psi_G}$ satisfy
the separability condition $\alpha(g_A\otimes g_B, h)\equiv \alpha(g_A\otimes g_B h_g)= \alpha_A(g_A) \alpha_B(g_B) \beta(h_g)$ for every $g\in G$, then it is possible to prove \cite{hiz2} that the
von Neumann entropy of the $G$-state corresponding to the bipartition
$(A,B)$ is:
$S(\ket{\Psi_G})= -\sum_{[h]\in G_{AB}} |N_A N_B \beta(h)|^2 \log_2 |N_A N_B \beta(h)|^2
\label{S}$, 
where $N_X^2:= \sum_{g_X\in G_X} |\alpha_X (g_X)|^2$, for $X= A,B$.
By convexity of $S$ we have $S(\ket{\Psi_G}) \leq
S(\ket{G}) = \log_2 |G_{AB}|$.

This formalism is remarkably well suited to describing topologically
ordered states. In many quantum spin systems, topological order arises
from a mechanism of closed string condensation and the group $G$ is
the group of closed strings on a lattice \cite{closedstrings}. An important example
of topologically ordered system is given by Kitaev's toric code, which
provides a model for which at zero temperature 
topological memory and topological quantum computation are robust
against arbitrary local perturbations \cite{kitaev}. The model is defined on a
square lattice with spin-$1/2$ degrees of freedom on the edges and
periodic boundary conditions. To every
plaquette $p$ we associate the operator product of $\sigma^x$ on all
the spins that comprise the boundary of $p$, i.e., $X_p = \prod_{j\in
  p} \sigma^z_j$. To every vertex $s$ we associate the product of $\sigma^z$ on all the spins
connected to $s$: $Z_s = \prod_{j\in s} \sigma^x_j$. The operators $X_p$ generate a group $G$ of closed
string-nets.
The Hamiltonian of the toric code in an external magnetic field is:
\be\label{toric}
H_{\rm toric} = - \sum_p X_p -\lambda \sum_s Z_s +(1-\lambda ) \sum_j \sigma ^z_j,
\ee
where we have introduced a control parameter $\lambda$. A second order quantum phase transition at $\lambda_c\sim
0.7$ separates a
spin-polarized phase ($0\leq \lambda < \lambda_c$) from a
topologically ordered phase ($ \lambda_c <\lambda \leq 1$) \cite{top-qpt,topqpt-num}. The ground state of
$H_{\rm toric}$ is a $G$-state throughout the entire topological phase. It is $G$-uniform at the toric code point
$\lambda=1$, and becomes less uniform as $\lambda$ decreases to
$\lambda_c$.

We now wish to argue that the separability condition for $\alpha(g)$
is satisfied throughout the entire
topological phase, and hence by convexity $S_\lambda \le S(\ket{G}) =
\log_2 |G_{AB}|$ for $\lambda_c < \lambda \leq 1$, with the bound 
saturated at the toric code point. At $\lambda=0$ the ground state is
the uniform superposition of closed strings. The $\lambda$ term in
Eq.~(\ref{toric}) is a tension for the strings. As we increase
$\lambda$, larger strings become less favored in the ground
state. Everywhere in the topological phase, that is, for
sufficiently small
$\lambda$, the ground 
state is still the superposition (with positive coefficients \cite{topqpt-num}) of closed strings $g\in G$. The expectation value $\langle g \rangle $ of any closed string $g
\in G$ of length $l$ (a Wilson loop) 
can be written as $\langle g \rangle = C_t ^2 e^{(1-\lambda ) l(g)}$, where $C_t$ is
a constant that does not depend on $g$ (due to translational
invariance). Similarly, in the
polarized phase we have $\langle g \rangle = C_p ^2 e^{-\lambda a(g)}$,
where $a$ is the area enclosed by the string \cite{Kogut}. Now, we
know that $\langle g \rangle = |\alpha(g)|^2$ at any point in the
topological phase, since the ground state is a $G$-state and does not
contain any open strings. Since the length $l$ for a given string $g=g_A\otimes g_B h_g$
can be decomposed as a sum of the corresponding substrings,
$l=l_A+l_B+l_{AB}$, we have $\alpha(g) = C_t e^{-l(g)/2} = C_t e^{-l_A/2}
e^{-l_B/2} e^{-l_{AB}/2} \equiv \alpha_A(g_A) \alpha_B(g_B) \beta(h_g)$,
i.e., we have separability.

\textit{Fractal boundary}.--- Henceforth we consider the toric code point $\lambda=1$, where
$S=\log_2 |G_{AB}|$. We define bipartitions by drawing strings
along the
edges of the lattice. One can prove \cite{hiz} that
$\log_2 |G_{AB}|$ is the number of independent plaquette operators $A_p$
acting on both subsystems $A$ and $B$, which in turn is the number of
squares that have at least one side adjacent to
the boundary $p$ of the region $\mathcal A$, see Fig.~\ref{duallattice}. How do we measure $p$?  
We shall show
that the support of the mixed part of the reduced density matrix is
given exclusively by
the spins on the boundary.
This mixed part is the only part contributing to the entanglement
between the $A$ and $B$ partitions.
Therefore we define the length $p$ as the
number of boundary spins.
Indeed, letting
$Q_X = | G_X|^{-1/2}\sum_{g_X\in G_X} g_X$, with
$X=A,B$, the ground state can be written as $\ket{G} =|
G_{AB}|^{-1/2} \sum_{h\in G_{AB}}h_A h_B Q_AQ_B\ket{0}$.
It follows from the definition of $G_{AB}$
that we can pick $h_A$ up to local transformations of the loops inside
 $A$ and $B$. Specifically, we can pick $h_A$ as
acting only on the spins on the boundary. Since $Q_A,Q_B$ are local
operators, the reduced density matrix of the $A$-subsystem is equivalent to one separable as
$\tr_B[\ket{G}\bra{G}] = \ket{\psi}\bra{\psi}\otimes\tilde{\rho}_A$,
where $\ket{\psi}$ is a pure state describing $A$'s bulk, while
the mixed part is
$ \tilde{\rho}_A =| G_{AB}|^{-1} \sum_{h\in G_{AB}}h_A\ket{0} \bra{0}
h_A $, where $h_A$
acts exclusively
on the spins along the boundary of $A$ \cite{fhhw}.
Thus $S/p$ is the average entanglement per spin in the support of $ \tilde{\rho}_A$.

%%%%%%%%%%%%%%%%%%%%%%
We now consider the case of a bipartition
defined by a closed fractal curve. Since the model studied here is
defined on a square lattice, we consider
bounded regions of $\mathbb{Z}^{2}$ depending on a parameter $n$,
denoted by $\mathcal{A}_{n}$. Here
$n$ represents the number of steps in the iteration generating the
fractal curve. The \emph{perimeter} of $\mathcal{A}_{n}$
is denoted by
$p\left(  \mathcal{A}_{n}\right)  $.
The number of squares of size one adjacent to the boundary of $\mathcal{A}_{n}$ is the entanglement $S(\mathcal{A}_{n})$ associated to the bipartition $(\mathcal{A}_{n}, \mathcal{B}_n)$. We are interested in the large $n$ limit of the ratio between entanglement and perimeter:
$\gamma\left(  \mathcal{A}\right):=
\lim_{n\rightarrow\infty}S(\mathcal{A}_{n})/p(\mathcal{A}_{n})$. One
might expect the scaling law $S = p -1$ to be independent of the
geometric properties of the bipartition, but this is not the
case. From Fig.~\ref{duallattice}, we see that 
when the boundary of $A$ has some inward angles, or wells, or other
``kinks'', the number of squares adjacent to it is less than the
length of the boundary around it. For instance, an inward angle, a
well, and a hole all have just one adjacent square of side $1$ but
they have lengths $2,3,4$ in the lattice spacing unit, respectively.
We call $\alpha$ and $h$ the number of inward angles and holes,
respectively. It is not hard to show that \cite{comment}
\be
\label{sp}
 S = p- \alpha -3 h.
\ee
We wish to study how these numbers scale for a fractal expansion, and find
the corresponding scaling of the entanglement.

 In the following, we shall compute $\gamma$ for several fractal
 curves. The results are summarized in Table~\ref{tab1}. The main
 result is that, depending on the fractal region, $\gamma$ can be a
 fractional number. The Hausdorff dimension $D$ of the fractal does
 not 
uniquely determine the value of $\gamma$, but (in all the examples considered) we have the bound
$\gamma\le D^{-1}$.

\ignore{
\begin{figure}
  \includegraphics[width=7cm]{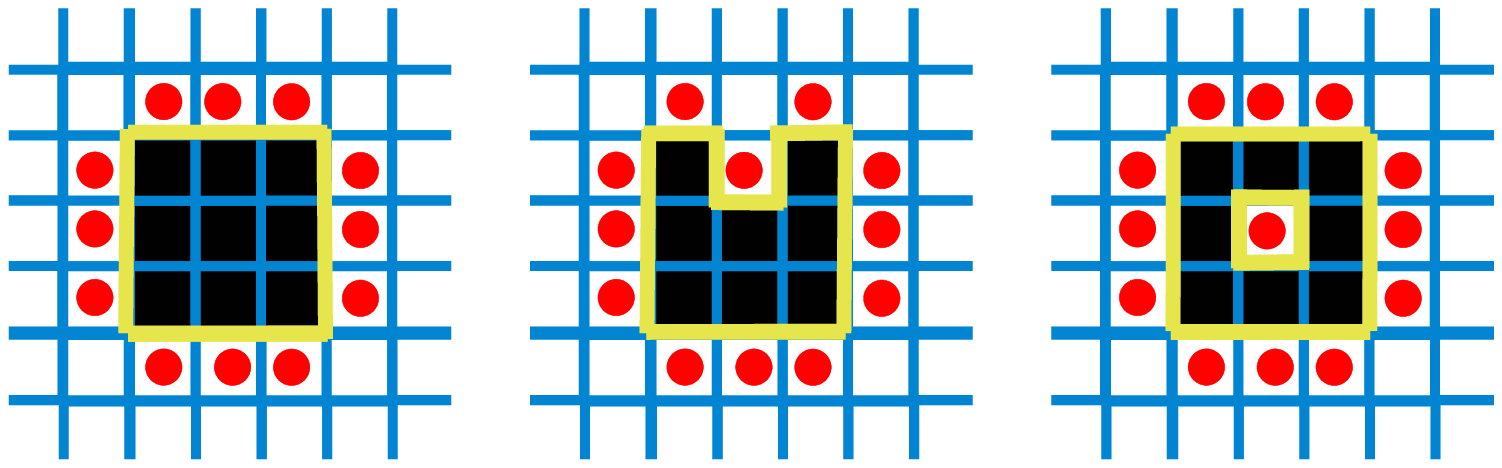}\\
  \caption{The drawings show different bipartitions of the
system. The subsystem $\mathcal A$ consists of all the spins marked
by the black squares. The entanglement is given by the number of
plaquette operators acting on both subsystems, marked by red dots. For a
regular figure (left), this number coincides with the perimeter
$p$, which is the number of spins along the boundary (in yellow). Every time there is an inward angle, there is one such operator
for three units of length. The well (middle) contains two inward
angles. A hole (right) of size $1$ accounts for $4$ units of length
and contains only one star operator.}
   \label{duallattice}
 \end{figure}  
 }

\begin{figure}[tbp]
\begin{equation*}
\leavevmode\hbox{\epsfxsize=7 cm
   \epsffile{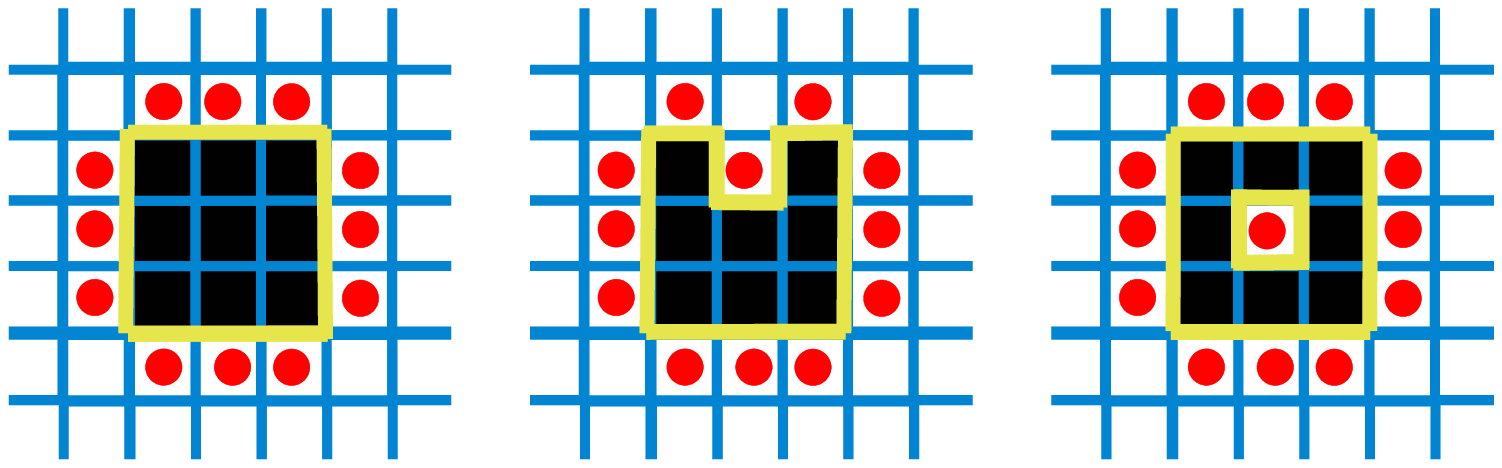}}
\end{equation*}%
\caption{The drawings show different bipartitions of the
system. The subsystem $\mathcal A$ consists of all the spins marked
by the black squares. The entanglement is given by the number of
plaquette operators acting on both subsystems, marked by red dots. For a
regular figure (left), this number coincides with the perimeter
$p$, which is the number of spins along the boundary (in yellow). Every time there is an inward angle, there is one such operator
for three units of length. The well (middle) contains two inward
angles. A hole (right) of size $1$ accounts for $4$ units of length
and contains only one star operator.} 
\label{duallattice}
\end{figure}

\textit{Examples}.--- The \emph{Sierpinski carpet} on $\mathbb{Z}^{2}$, denoted by $\mathcal{S}_{n}%
$, is a bounded region of $\mathbb{Z}^{2}$ defined iteratively in the
following way: \emph{(i)} $\mathcal{S}_{1}$ is a $3\times3$ square without the central $1\times1$
square. The Sierpinski carpet $\mathcal{S}_{1}$ has a single square
\emph{hole}. \emph{(ii)} $\mathcal{S}_{n+1}$ is a bounded region inscribed on a $3^{n}\times
3^{n}$ square on $\mathbb{Z}^{2}$. This is obtained by placing $8$ copies of $\mathcal{S}_{n}$ on all quadrants of the square, but the central one (see Fig.~\ref{figure2}). Given the recursive structure of $\mathcal{S}_{n}$, direct calculations show that
$\alpha(\mathcal{S}_{n})=\frac{1}{14}8^{n}-\frac{4}{7}$. The
number of equal holes of side $3^{i}$ is $8^{n-1-i}$, so $h(n) =8^{n-1} $. Observe that the external perimeter of
$\mathcal{S}_{n}$ is $4\times3^{n}$. Then the perimeter $p(n)$ is
$p(\mathcal{S}_{n})   =4(  3^{n}+3^{n-1}+\sum\nolimits_{i=0}^{n-2}(
3^{i}\times8^{n-1-i}))   =4\left(  4\times3^{n}+8^{n}\right)  /5$. With this information, from Eq.~(\ref{sp}) we obtain $\gamma\left(  \mathcal{S}_{n}\right)  =99/224$.

\ignore{
\begin{figure}
  \includegraphics[width=7cm]{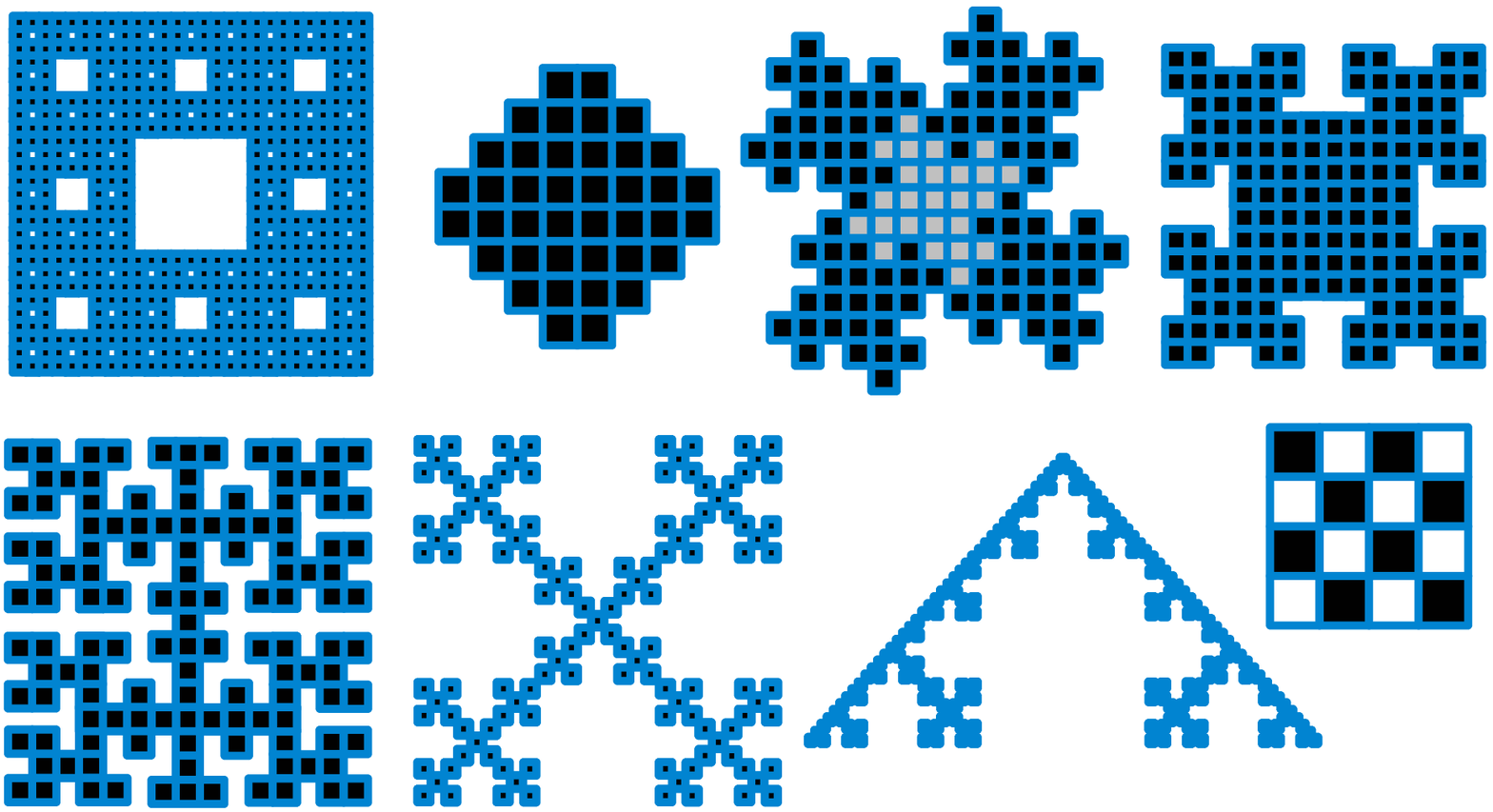}\\
  \caption{(Color online) Top, left to right: Sierpinski carpet $\mathcal{S}_{3}$, Greek cross
  $\mathcal{G}_{3}$, Minkowski sausage $\mathcal{I}_{3}$,  T-square
  $\mathcal{E}_4$. Bottom, left to right: Moore polygons $\mathcal{M}_{3}$,
  Vicsek fractal $\mathcal{V}_{3}$, half perimeter of the Koch polygon
  $\mathcal{K}_5$, $4\times 4$ chessboard $\mathcal C_4$.}
   \label{figure2}
 \end{figure}
 }

\begin{figure}[tbp]
\begin{equation*}
\leavevmode\hbox{\epsfxsize=7 cm
   \epsffile{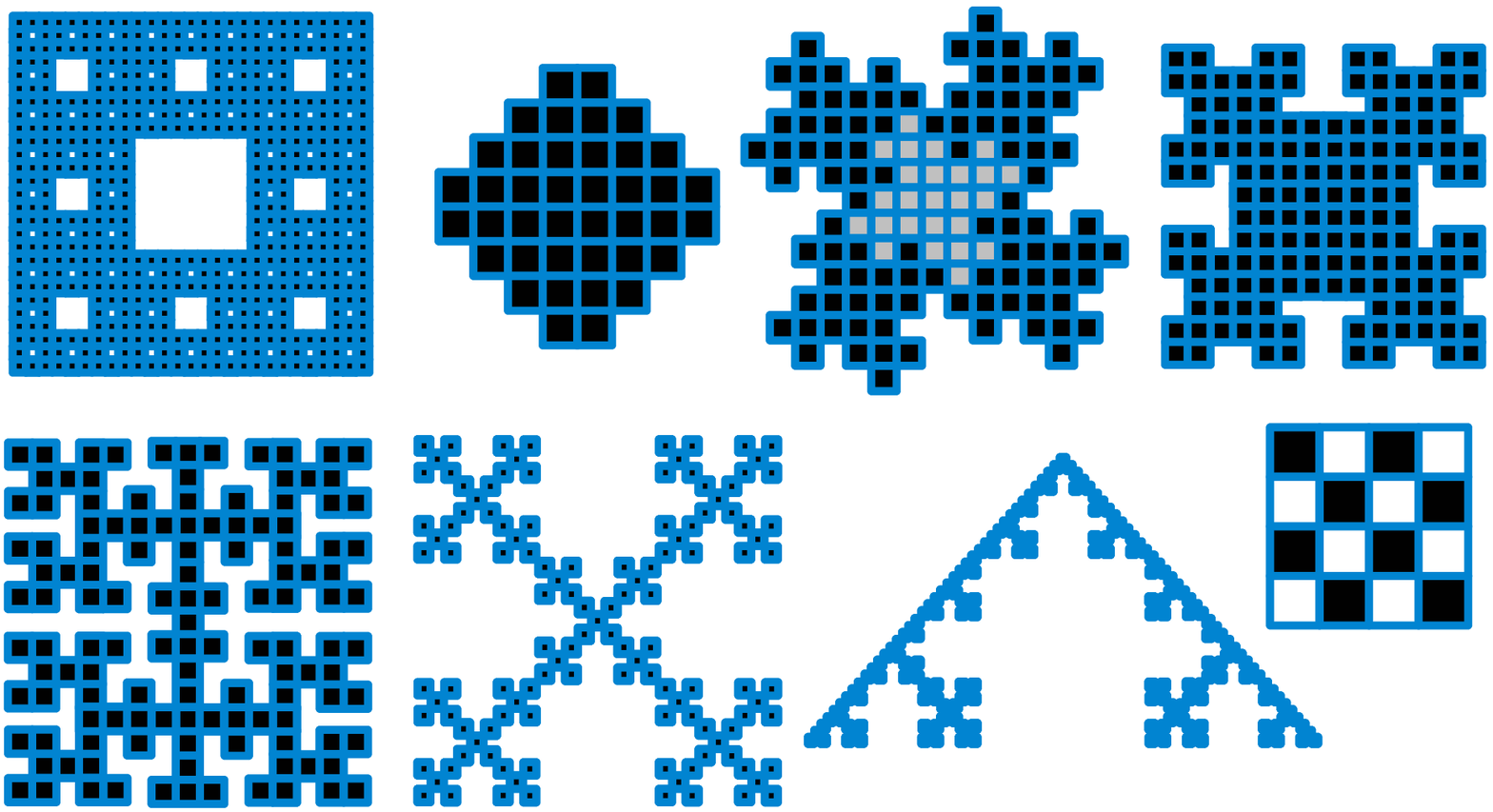}}
\end{equation*}%
\caption{(Color online) Top, left to right: Sierpinski carpet $\mathcal{S}_{3}$, Greek cross
  $\mathcal{G}_{3}$, Minkowski sausage $\mathcal{I}_{3}$,  T-square
  $\mathcal{E}_4$. Bottom, left to right: Moore polygons $\mathcal{M}_{3}$,
  Vicsek fractal $\mathcal{V}_{3}$, half perimeter of the Koch polygon
  $\mathcal{K}_5$, $4\times 4$ chessboard $\mathcal C_4$.} 
\label{figure2}
\end{figure}

The \emph{Greek cross on} $\mathbb{Z}^{2}$,
denoted by $\mathcal{G}_{n}$, is a polygon in $\mathbb{Z}^{2}$ defined
by a closed path of length $p\left(  \mathcal{G}_{n}\right)  =8n+8$, including
the point $\left(  0,n\right)$ and the step $\left\{
(0,n),(1,n)\right\}$. The path maximizes the number of inward angles
over all the closed paths of the same length including the point
$\left(  0,n\right)$. Fig.~\ref{figure2} gives the first few
instances. 
It is then evident that $\alpha(\mathcal{G}_{n})=4n$.
For this polygon, $h(n)=0$ and thus from Eq.~(\ref{sp}) we have $S(n) = p(n) -\alpha(n)$. Therefore,
$\gamma\left(  \mathcal{G}_{n}\right)  =1/2$.

The \emph{Minkowski sausage} $\mathcal{I}_{n}$ is a
polygon in $\mathbb{Z}^{2}$ defined as follows: \emph{(i)}
$\mathcal{I}_{0}$ is a square of side one. \emph{(ii)} 
$\mathcal{I}_{n+1}$ is obtained by replacing each side of $\mathcal{I}_{n}$ by
a path of length three. The angles in the path are determined by the position
of the side in $\mathcal{I}_{n}$. The first and third segments of the path
follow the direction of the replaced side. The two angles are first left then
right. Analogously, we can construct $\mathcal{I}_{n+1}$ by attaching to the
sides of $\mathcal{I}_{n}$ four of its copies (see Fig.~\ref{figure2}). The polygon
$\mathcal{I}_{n}$ can be used to tessellate the plane. From the
definition, we can determine $p\left(  \mathcal{I}_{n}\right)  =4\times3^{n}$ and $\alpha\left(
\mathcal{I}_{n}\right)  =2\times3^{n}-2$. Here too we have $S(n) =
p(n) -\alpha(n)$. Hence, $\gamma\left(  \mathcal{I}_{n}\right)
=1/2$. 

The \emph{Moore polygon} $\mathcal{M}_{n}$ is a \textquotedblleft closed
version\textquotedblright\ of the Moore curve. It is a polygon in $\mathbb{Z}^{2}$ defined
by a closed path expressed as an $L$-system. A
\emph{Lindenmayer system} (for short, $L$\emph{-system}) \cite{rs80} is a
quadruple $\left\langle V,C,A,R\right\rangle $, where $V$ is a set of
\emph{variables}, $C$ a set of \emph{constants}, $A$ a set of \emph{axioms},
and $R$ a set of \emph{production rules}. An $L$-system allows the recursive
construction of words (or, equivalently, sequence of symbols) whose letters
are elements from $V$ and $C$. An axiom is a word at time $t=0$. At each time
step $t+1$, the production rules are applied to the word given by the
$L$-system at time $t$. Only variables are replaced according to the
production rules. On the basis of these definitions, we can write
$\mathcal{M}_{n}=\left\langle V,C,A,R\right\rangle $, where $V=\{a,b\}$,
$C=\{+,-\}$, $A=\{aFa+F+aFa\}$, and $R=\{a\rightarrow-bF+aFa+Fb-,b\rightarrow
+aF-bFb-Fa+\}$. The letter $F$ indicates a segment of length one in
$\mathbb{Z}^{2}$. The first segment of $\mathcal{M}_{0}$ specified by the
axiom in $A$ is $\{(0,0),(1,0)\}$. The symbols $+$ and $-$ mean
\textquotedblleft turn left in $\mathbb{Z}^{2}$\textquotedblright\ and
\textquotedblleft turn right in $\mathbb{Z}^{2}$\textquotedblright,
respectively. The sequences $-+$ and $+-$ have no meaning and can be deleted.
For instance, the polygon $\mathcal{M}_{1}$ is then given by the the following
word: $-bF+aFa+Fb-F-bF+aFa+FbFbF+aFa+Fb-F-bF+aFa+Fb-F$. Notice that in order
to close $\mathcal{M}_{1}$ we need to replace $\cdots+Fb-F$ with $\cdots+FbF$
in the obtained word. This operation is required for every $n$. Once we have
generated the polygon, we blow it up by replacing each square of side one with a square
comprising four of its copies.
The occurrences of letter $F$ in the word produced by $\mathcal{M}_{1}$ is
$16$. In general, the number of occurrences of $F$ in the word produced by
$\mathcal{M}_{n}$ equals the perimeter of $\mathcal{M}_{n}$. From the
definition, this is $p(\mathcal{M}_{n})=2\times4^{n+1}$, taking into account
the blowing up operation. The number of $-$ (``turn right'') symbols, excluding the initial
one, in the word produced by $\mathcal{M}_{n}$, is exactly equal to the number
of inward angles of $\mathcal{M}_{n}$: $\alpha(\mathcal{M}_{n})=\frac{2}{5}\left(
-1\right)  ^{n}+\frac{8}{5}4^{n}-2$. From $S =
p(\mathcal{M}_{n})-\alpha(\mathcal{M}_{n})$, we can compute
$\gamma\left(  \mathcal{M}_{n}\right) =4/5$.

The \emph{Vicsek snowflake on} $\mathbb{Z}^{2}$, denoted by $\mathcal{V}_{n}$, is a bounded region of $\mathbb{Z}^{2}
$ defined
iteratively
as follows: \emph{(i)} $\mathcal{V}_{0}$ is a single $1\times 1$ square. \emph{(ii)} We obtain $\mathcal{V}_{n+1}$ by attaching $4$
copies of $\mathcal{V}_{n}$ to its corners (see Fig. \ref{figure2}). Each square comprising $\mathcal{V}_{n}$ has side one.
For this fractal we have $p\left( \mathcal{V}_{n}\right) =20\times
5^{n-1}$ and $\alpha \left( \mathcal{V}_{n}\right) =2\times 5^{n}-2$. The number of adjacent squares is
$S(n) = p(n) -\alpha(n)$, which gives  $\gamma (\mathcal V)=\frac{1}{2}$.

The \emph{quadratic Koch polygon}, $\mathcal{K}_{n}$, is a polygon in $%
\mathbb{Z}^{2}$ based on the Koch curve. Essentially, it consists of a
region bounded by two mirroring copies of the Koch curve. As the Moore
polygon, $\mathcal{K}_{n}$ is defined by an $L$-system and specified by a
path. The path giving rise to $\mathcal{K}_{0}$ is given axiomatically as $%
\{(0,0),(1,0)\}$. Then $\mathcal{K}_{0}$ is a square of side one. The
production rule is $F\rightarrow F+F-F-F+F$, where $F$ indicates again a
segment of length one in $\mathbb{Z}^{2}$. The fractal has a pattern similar to that of the Vicsek snowflake and indeed has the same Hausdorff dimension
(see Table~\ref{tab1}).
Nevertheless, the results for the scaling of the entanglement are different. The perimeter can be computed as $p(n) = 4\times 5^n$. The number $h(n)$ of holes is $h= \frac{18}{125}\times 5^n +\frac{1}{3}3^n-1$, for $n\ge3$. One can easily see that $\alpha = (p-4h)/2$ and therefore from Eq.~(\ref{sp}) $S=\frac{p}{2}-h = \frac{232}{125} 5^n  -\frac{1}{3}3^n+1$. In the limit of large $n$, we obtain $\gamma = 58/125$.

The \emph{T-square polygon on }$\mathbb{Z}^{2}$,
$\mathcal{E}_{n}$, is obtained by superimposing four copies of $\mathcal{E}%
_{n-1}$ on the corners of a square of side $2^{n+1}$. The area covered by each
copy is exactly a square of side $2^{n}$. The perimeter of $\mathcal{E}_{n}$ is
$p\left(  \mathcal{E}_{n}\right)  =16\times3^{n}-8\times2^{n}$.
We have $S\left(  \mathcal{E}_{0}\right)  =4$, $S\left(  \mathcal{E}%
_{1}\right)  =24$, and
$S\left(  \mathcal{E}_{n}\right)    =3S\left(  \mathcal{E}_{n-1}\right)
+2^{n+1}-8 =\frac{80}{9}3^{n}+2^{n+1}-8+24\times S(n,3) =\frac{92}{9}3^{n}-4\times2^{n}+4,
$
where $\mathcal S(n,3):=(1+3^{n-2}-2^{n-1})/2$ is the $n$-th Stirling number of the
second kind. Hence, $\gamma=1/2$.

\emph{The chessboard} $\mathcal C_n$ is the bounded region of $\mathbb{Z}^{2}$ defined as follows. Let $\mathcal C_1$ be a $2\times 2$ square with two holes in the upper right and bottom left corner. Then $\mathcal C_{n+1}$ is obtained by placing $4$ copies of $\mathcal C_n$ on all the quadrants of a $2^n \times 2^n$ square on $\mathbb{Z}^{2}$. The perimeter is $p = 2n$. The number of adjacent squares is exactly $h=n/2$. Therefore it is immediate that $\gamma = N_s/p = 1/4$ for every size $n$. It is obvious that this is a lower bound for the entanglement on the square lattice for a state in $\mathcal L$, since the chessboard maximizes the number of holes of side $1$.
\begin{table}[t]
\caption{Fractal entanglement $\gamma$, perimeter $p(n)$, entropy of
  entanglement $S(n)$ for a state in $\mathcal L$ for several fractal
  bipartitions $(A,B)$ of the square lattice. Here $D$ is the
  Hausdorff dimension of the curve separating the regions ${\mathcal
  A}_n$ and ${\mathcal B}_n$. For $p(n)$ and $S(n)$ only the leading term is shown.}
%\begin{ruledtabular}
\begin{tabular}{l|cccc}
  Fractal& $\gamma$ & $p(n)$ & $S(n)$ &$D$ \\
\hline
1. Sierpinski carpet & $\frac{99}{224}$ & $\frac{4}{5}8^n $& $\frac{99}{280} 8^n$&$\frac{\log 8}{\log 3}$ \\
2. Greek Cross & $\frac{1}{2}$ &$8n$& $4n$& $2$ \\
3. Minkowski Sausage  & $\frac{1}{2}$ &$4\times 3^n$&$2\times 3^n $& $\frac{\log 5}{\log 3}$\\
4. Vicsek Snowflake & $\frac {1}{2}$ &$4\times 5^{n}$ & $2\times 5^n $ & $\frac{\log 5}{\log 3}$\\
5. Quadratic Koch & $\frac{58}{125}$&$4\times 5^n$ & $\frac{232}{125} 5^n  $&$\frac{\log 5}{\log 3}$\\
6. Moore Polygon & $\frac{4}{5}$ & $2\times 4^{n+1}$ & $\frac{32}{5} 4^n $& $\frac{\log 9}{\log 6}$ \\
7.  T-Square &$\frac{1}{2}$ & $16\times 3^n$& $\frac{92}{9}3^{n}$&$2$ \\
8. Chessboard & $\frac{1}{4}$ &$8n^2$ & $2n^2$ & $2$ \\
\end{tabular}
%\end{ruledtabular}
\label{tab1}
\end{table}

%%%%%%%%%%%%%%%%%%%%%%
\textit{Conclusions}.---This work has, for the first time,
explored the relationship between entanglement entropy and the
fractality of the bipartition in a spin system. We have calculated the scaling of
entanglement $S$ with the length $p$ of the boundary in the ground
state of the $Z_2$ topological phase associated with the toric code,
for various fractal boundaries. We have shown that this provides
an upper bound on the entanglement in the entire topological phase. Unlike
the case of a regular boundary, the ratio $\gamma=S/p$ for large $p$
is not exactly $1$ but a smaller fraction, so that
the general bound for the area law is still obeyed. The fractal nature
of the bipartition is revealed in the total amount of entanglement
present in the system. There is less entanglement in a fractal
bipartition. We also found that the ratio $\gamma$ is always at most
the inverse of the Hausdorff dimension $D$. We conjecture this last
claim to hold in general for topologically ordered states. Moreover, different fractals with the same Hausdorff
dimension can have different $\gamma$, so that this is a useful
quantity to classify fractals with.
%%In this work, we have considered
We chose the toric code because in this case it is simple to compute the
entanglement. It would be interesting to consider other types of
topologically ordered states and explore
whether the behavior we have observed is general for any quantum
system with finite correlation length. Finally, since the scaling of entanglement with the boundary of the system is less than $1$, we
believe that a renormalization group algorithm based on blocks of
spins that grow like fractals, might be potentially more
efficient. Indeed, in this regard the chessboard appears to be the
most attractive of all the fractals we have considered.

%%%%%%%%%%%%%%%%%%%%%%

{\it Acknowledgments}.---Research at Perimeter Institute for
Theoretical Physics is supported in part by the Government of Canada
through NSERC and by the Province of Ontario through MRI.  D.A.L.'s
work was supported by NSF under grants No. CCF-726439,
No. PHY-802678 and No. PHY-803304. D.A.L. acknowledges the hospitality
of IQI-Caltech where part of this work was performed. Research at IQC is supported by
DTOARO, ORDCF, CFI, CIFAR, and MITACS.


\begin{thebibliography}{99}

\bibitem{ent-review}
L.~Amico, R.~Fazio, A.~Osterloh, and V.~Vedral, Rev. Mod. Phys. {\bf 80}, 517 (2008).

\bibitem{entqpt}
A. Osterloh, L. Amico, G. Falci, R. Fazio, Nature {\bf 416}, 608 (2002);
T.J. Osborne and M.A. Nielsen, Phys. Rev. A {\bf 66}, 032110 (2002);
G. Vidal, J.I. Latorre, E. Rico, A. Kitaev, Phys. Rev. Lett. {\bf  90}, 227902 (2003);
F. Verstraete, M. Popp, J. I. Cirac, Phys. Rev. Lett. {\bf  92}, 027901 (2004);
L.-A. Wu, M.S. Sarandy, D.A. Lidar, Phys. Rev. Lett. {\bf 93}, 250404 (2004).

\bibitem{hiz}
A. Hamma, R. Ionicioiu, P. Zanardi, Phys. Lett. A {\bf 337}, 22 (2005);
{\it ibid.}, Phys. Rev. A {\bf 71}, 022315 (2005)

\bibitem{topentropy}
A. Kitaev, J. Preskill, Phys. Rev. Lett. \textbf{96}, 110404 (2006);
M. Levin, X.-G. Wen, Phys. Rev. Lett. \textbf{96}, 110405 (2006).

\bibitem{vidal}
G. Vidal, Phys. Rev. Lett. {\bf 99}, 220405 (2007).

\bibitem{hastings-gap}
M.B. Hastings, T. Koma, Comm. Math. Phys. {\bf 265}, 781 (2006).

\bibitem{arealawrev}
J. Eisert, M. Cramer, M.B. Plenio, eprint arXiv:0808.3773.

\bibitem{arealaw}
M.M. Wolf, F. Verstraete, M.B. Hastings, J.I. Cirac,
Phys. Rev. Lett. {\bf 100}, 070502 (2008).

\bibitem{kitaev}
A.Y. Kitaev, Annals of Phys. {\bf 303}, 2 (2003).

\bibitem{topqpt-num}
A. Hamma, W. Zhang, S. Haas, D. Lidar, Phys. Rev. B {\bf 77}, 155111
(2008).

\bibitem{Kogut}
B. Kogut, Rev. Mod. Phys. {\bf 51}, 659 (1979).
  
\bibitem{lidarfractal}
O. Malcai, D.A. Lidar, O, Biham, D. Avnir, Phys. Rev. E {\bf 56}, 2817
(1997). 

\bibitem{fractalbook}
M. Barnsley, \emph{Fractals everywhere} (Academic Press, New York, 1988).

\bibitem{closedstrings}
{X.-G. Wen}, \emph{Quantum Field Theory of Many-Body Systems}, (Oxford
Univ. Press, Oxford, 2004).

\bibitem{top-qpt}
A. Hamma, D.A. Lidar, Phys. Rev. Lett. {\bf 100}, 030502 (2008); 
S. Trebst \textit{et al}., Phys. Rev. Lett. \textbf{98}, 070602 (2007).

\bibitem{hiz2}
A. Hamma, R. Ionicioiu, P. Zanardi, Phys. Rev. A {\bf 72}, 012324 (2005).
\bibitem{fhhw}
%%D.~Flammia,
S.T.~Flammia, A.~Hamma, T.L.~Hughes, X.-G.~Wen, Phys. Rev. Lett. {\bf 103}, 261601 (2009).

\bibitem{comment}
The entanglement $S$ is the number of squares in the
dual lattice that have at least one side adjacent to the boundary
of the region ${\cal A}$. For a figure that is a square of perimeter
$L$ with a $1 \times 1$ hole in the bulk, the total perimeter is $p
= L+4$. The number of adjacent squares is $S = L+1$ because there
are $L$ adjacent squares on the external boundary, and one
inside. Thus $S = p - 3$. With $h$ holes we have $p = L+4h$ and $S =
L+h$, so that $S = p-3h$. A similar counting argument which accounts
for inward angles leads to Eq.~(\ref{sp}).

\bibitem{fermions} D.~Gioev, and I.~Klich, Phys. Rev. Lett. {\bf 96}, 100503 (2006).

\bibitem {rs80}
G. Rozenberg and A. Salomaa, \emph{The mathematical theory of L
systems} (Academic Press, New York, 1980).
\end{thebibliography}
\end{document}